\documentclass[a4paper,11pt]{article}
\usepackage{pos}

\usepackage{amsmath}
\usepackage{amsfonts}
\usepackage{bbold}
\usepackage{color}
\usepackage{graphicx}
\usepackage{subcaption}
\usepackage{algpseudocode}
\usepackage{hyperref}
\usepackage{hhline}
\usepackage{comment}

\title{The Fuzzy Onion: An Initial Study}

\author*[a,b]{S. Kov\'{a}\v{c}ik}
\author[a]{J. Tekel}
\author[a]{M. Hrmo}

\affiliation[a]{Department of Theoretical Physics, Faculty of Mathematics, Physics and Informatics, \\
Comenius University in Bratislava, \\
Mlynsk\'a dolina, 842 48, Bratislava, Slovakia}
\affiliation[b]{Department of Theoretical Physics and Astrophysics, Faculty of Science, Masaryk University, \\
Brno, Czech Republic}

\emailAdd{samuel.kovacik@fmph.uniba.sk} 

\abstract{In our previous contribution, we introduced a matrix formulation of a three-dimensional quantum space named the fuzzy onion. The novel part of the construction is the radial derivative term, which has been defined to recover the correct continuum limit. Here, we describe a numerical simulation of the scalar field theory in this space and test some physical properties of the model with emphasis on the interaction between neighbouring layers.}

\FullConference{%
Corfu Summer Institute 2023 "School and Workshops on Elementary Particle Physics and Gravity",\\
23 April - 6 May, and 27 August - 1 October, 2023\\
Corfu, Greece}

\begin{document}
\maketitle

\section{Introduction}

A long-term goal of physics is to understand the phenomenon of quantum space to see if consistent physical theories can be built on it and then to draw verifiable predictions from that theory. One of the most studied examples is the fuzzy sphere, which has been studied in various contexts \cite{Hoppe,Madore:1991bw,Kovacik:2013vbk, Galikova:2013zca, Vitale:2012dz,Hammou:2001cc,Steinacker:2011ix,Gubser:2000cd, Szabo:2001kg, Bagger:2007vi,Skenderis:2008qn,Baez:1998he,Kovacik:2018thy,Ydri:2014rea, Panero:2006bx,GarciaFlores:2009hf}. 

The upside of the fuzzy sphere model is its simplicity and transparency, which is the reason behind the numerous studies performed on it. The downside is obvious ––– the corresponding geometry differs from the space we live in. Had we been living in a two-dimensional space, the fuzzy sphere would be a good candidate for the model of quantum space. Instead, it is either an interesting mathematical model to study basic properties of quantum space or a useful regularisation tool, \cite{Kuroki:1998rx,Zhu:2022gjc}. In our previous contribution, we proposed a way of connecting fuzzy spherical layers to form the fuzzy onion space. This model has two advantages: it is still nearly as simple as the original fuzzy sphere model, and it describes a three-dimensional space.

The construction of the fuzzy onion as presented there, had two nontrivial steps. The first was connecting various layers. If the fundamental length scale is fixed along increasing sphere radius, one has to use matrices of increasing size. Comparing them, for example, to calculate the radial derivative is naively impossible. This has been overcome by using the Fourier transform and discarding or adding the unmappable degrees of freedom. The second step was to define the full Laplace operator that captures the geometry of the space. There is some ambiguity in this definition; ours differs from that of \cite{seik, patricia}. Our goal was to reproduce the results of the model of three-dimensional quantum space studied in \cite{Kovacik:2013vbk, Galikova:2013zca} that has been studied thoroughly and has well-established commutative limit; some preliminary tests of the construction have been passed, see \cite{Kovacik:2023zab}. 

\section{The fuzzy sphere and the fuzzy onion}
We begin with a brief reminder of the fuzzy sphere and the fuzzy onion constructions. Fields on a sphere can be expanded into spherical harmonics, which form an infinite-dimensional representation of the $SO(3)$ generators $\hat{L}_i$:
\begin{eqnarray} \nonumber
 f(\theta,\varphi) &=& \sum_{lm} c_{lm} Y_{lm} (\theta,\varphi), \\
 \hat{{\cal{L}}} \ Y_{lm} &=& \hat{L}_i \hat{L}_i \ Y_{lm} = l (l+1) Y_{lm} , \\ \nonumber
 \hat{L}_3 \ Y_{lm}&=& m\ Y_{lm} ,
\end{eqnarray}
Finite-dimensional representations with momentum cut-off exist; that is, they have a maximum value of $l$. In this case, the fields can be represented as $N \times N$ matrices $Y_{lm}$, where $l=N-1$ is the maximal angular momentum. Fields are encoded into $N \times N$ matrices, and the angular momentum eigenstates now form the basis for Hermitian matrices of this size.

The matrix algebra generated by $Y_{lm}$ is closed under multiplication and noncommutative. Both fields on the sphere and Hermitian matrices can form an $so(3)$ representation space, but they differ in the momentum cut-off. The expansion coefficients serve as a map between functions on the sphere and Hermitian matrices. If one imposes a momentum cut-off on the sphere, this map can be bijective. However, this comes with a loss --- with a limited number of expansion terms, one can only approximate the Dirac $\delta$-function, or in other words, the spatial resolution is limited. In this way, Hermitian matrices can be used to encode fields on a sphere with limited spatial resolution. The minimal distinguishable scale is given by the parameter $\lambda$, where $[\hat{x}_i,\hat{x}_j] = 2 i \lambda \varepsilon_{ijk} \hat{x}_k$ and $\hat{x}_i$ are position operators. 

The radius of the fuzzy sphere is $r \sim N \lambda$, which means that if the minimal length scale is fixed, to describe a larger sphere, one needs a larger matrix. Layers of fuzzy spheres with increasing radius, denoted by the superscript ${}^{(i)}$, can then be put together into a larger matrix of the form
\begin{equation} \label{psi}
 \Psi = \begin{pmatrix}
\Phi^{(1)} & & &\\
 & \Phi^{(2)} & &\\
 & & \ddots &\\
 & & & \Phi^{(N_m)}
\end{pmatrix} ,
\end{equation}
where the angular part of the Laplace operator acts on each of the layers separately
\begin{equation} \label{L}
 {\cal{L}} \Psi = \begin{pmatrix}
 \hat{{\cal{L}}}^{(1)}\Phi^{(1)} & & &\\
 & \hat{{\cal{L}}}^{(2)}\Phi^{(2)} & &\\
 & & \ddots &\\
 & & & \hat{{\cal{L}}}^{(N_m)}\Phi^{(N_m)}.
\end{pmatrix}.
\end{equation}
To define the radial part of the Laplace operator, first, we need to define an operator that connects submatrices $\Phi^{(i)}$ on consecutive layers:
\begin{eqnarray}
 {\cal{U}} \Phi^{(i)} &=& \Phi^{(i+1)} \in {\cal{H}}(N+1), \\
 {\cal{D}} \Phi^{(i)} &=& \Phi^{(i-1)} \in {\cal{H}}(N-1), 
\end{eqnarray}
where ${\cal{H}}(N)$ is the set of Hermitian matrices of size $N$. This can be done by first doing the Fourier expansion and then discarding the unmappable coefficients when going one layer down or adding zeroes when going one layer up:
\begin{eqnarray} \nonumber
 \Phi^{(N)} &=& \sum \limits_{l=1}^{N-1} \sum \limits_{m=-l}^l c^{(N)}_{lm} Y_{lm}^{(N)}, \\ \nonumber
 &&\\ \nonumber
 &{\cal{D}} \uparrow \hspace{0.5cm} {\cal{U}} \downarrow &\\ \nonumber
 \Phi^{(N+1)} &=& \sum \limits_{l=1}^{N-1} \sum \limits_{m=-l}^l c^{(N+1)}_{lm} Y_{lm}^{(N+1)}, \\ \nonumber
 && \\ \nonumber
 &\text{ where }& \\ \nonumber
 c_{l,m}^{(N)} &=& c_{l,m}^{(N+1)} \text{ for: } l\le N-1,
 \\ \nonumber
 c_{N,m}^{(N+1)} &=& 0. 
\end{eqnarray} 
Using this, we are in a position to define the radial derivative and the radial part of the Laplace operator. This is the source of some ambiguity as infinitely many definitions of the derivative coincide in the $\lambda \rightarrow 0$ limit. We chose a simple option that reproduces the ordinary energy spectrum, but a more detailed study is needed here:
\begin{equation} \label{dr}
 \partial_r \Phi^{(N)} = \frac{{\cal{D}}\phi^{(N+1)} - {{\cal{U}}}\phi^{(N-1)} }{2\lambda},
\end{equation}
and 
\begin{equation} \label{ddr}
 \partial^2_r \Phi^{(N)} = \frac{{\cal{D}}\phi^{(N+1)} -2\phi^{(N)} + {{\cal{U}}}\phi^{(N-1)} }{\lambda^2}.
\end{equation}

Using those, we can define the full Laplace operator that acts on the  $\Psi$ that describe fields that exist on a series of concentric fuzzy spheres of increasing radius, that is, on the fuzzy onion space:
\begin{equation}
 \Delta \Psi = \left(\frac{1}{r^2} \frac{\partial}{\partial_r} \left( r^2 \frac{\partial}{\partial r} \right) - \frac{{\cal{L}}^2}{ r^2} \right) \Psi
\end{equation}
It is trivial to define a potential on the fuzzy sphere using polynomials
\begin{equation} \label{fuzzy sphere potential}
V(\phi) = \mbox{Tr}\ P(\Phi),
\end{equation}
where the trace operation replaces the integration --- the same can be done on the fuzzy onion. Still, we must include $4\pi r$ as an integration measure as it differs across various fuzzy spheres. With this definition, there is something disconcerting. While the fields are smeared on each layer due to nonlocality, they are not smeared across various layers. This can be changed artificially by a smearing procedure
\begin{equation}
{\cal{S}} \phi^{(n)} = \frac{\phi^{(n)}+ \sum \limits_{i} \alpha_i \left({\cal{U}}^i \phi^{(n-i)} + {\cal{D}}^i \phi^{(n+i)}\right)}{1 + \sum \limits_{i} \alpha_i},
\end{equation}
for example with $\alpha_1 = \frac{1}{2}, \alpha_{2+}=0$:
\begin{equation}
{\cal{S}} \phi^{(n)} = \frac{ \phi^{(n)} + \frac{1}{2} {\cal{D}}\phi^{(n+1)} + \frac{1}{2} {\cal{U}}\phi^{(n-1)}}{2},
\end{equation}
and with it, we can take a potential that is nonlocal in each direction as
\begin{equation}
V(\Psi) = \sum \limits_{j=1}^{N_m} V \left({\cal{S}} \phi^{(j)}\right).
\end{equation}
The other option is to consider the model for strings stretching between various layers as in \cite{Steinacker:2022kji}.

\section{Scalar field theory}
The scalar field theory is defined by the action
\begin{equation} \label{action}
    S[\Psi] = 4\pi \lambda^2 \mbox{Tr } r\left( a\ \Psi {\cal{K}} \Psi + b \ \Psi^2 + c\ \Psi^4 \right), \ \ {\cal{K}} =  - \frac{1}{2} \Delta,
\end{equation}
where $r$ is a diagonal matrix that multiplies each layer with the corresponding radius. The mean values of observables are defined as:
\begin{align}
    \left\langle {\cal {O}}(\Psi)\right\rangle=\frac{1}{Z} \int d\Psi e^{-S(\Psi)}{\cal {O}}(\Psi),\ d\Psi=\prod_{N=1}^{M}d\Phi^{(N)}\ .\label{expectation}
\end{align}
Basically, this is a Hermitian matrix model in which we consider only a subset of matrices of a certain type and have a specific action that serves to define the integration measure. 

Hermitian matrices are partially described by their eigenvalues, and for a pure potential matrix model, the action depends only on them. The diagonalisation procedure and integrating the rest of the degrees of freedom produces a new term in the integration measure that can be interpreted as a logarithmic repulsion between the eigenvalues \cite{Tekel:2015uza}. In the case of the field theories on fuzzy spaces, the action involves a kinetic term that prevents diagonalisation, but analysing the behaviour of the eigenvalues still proves to be useful for understanding the model. 

Usually, one takes $a=1$ and then produces the phase diagram of the model with respect to $b,c$. While $c$ is positive, $b$ can be positive, zero or negative. For sufficiently large values of $b$, all eigenvalues reside close to the origin; this is called the disordered phase. There are two options for sufficiently negative values of $b$. There are two potential minima and either all eigenvalues are in one of them, this is called the uniformly ordered phase, or are split between them due to repulsion; this is the nonuniformly ordered phase. 

As was shown in great detail in \cite{Kovacik:2018thy}, the phase diagram for the scalar field theory on the fuzzy sphere converges when expressed in rescaled parameters
\begin{equation}
    \tilde{b} =\frac{b}{a N^{3/2}} , \ \tilde{c} = \frac{c}{a N^2}.
\end{equation}

This is interesting because, in our case, the value of $N$ varies across layers. That means that for fixed values of $b$ and $c$, different layers might prefer different phases. The assumption in this regard is that the outermost layer would set the behaviour as it has the largest number of degrees of freedom.  

To understand the behaviour of the model, we have performed Hamiltonian Monte Carlo (HMC) studies of the model. While the fuzzy sphere model has been studied with $N > 100$, the fuzzy onion model is more computationally demanding as one has to perform the Fourier transformation at each step. Since this is an initial study, we chose $N=10$ for the largest matrix. 

There were two questions we wanted to answer. Does the radial part of the Laplace operator work as intended and cause interaction between consecutive layers? Does the outermost layer dictate the behaviour when there is a mismatch between preferred phases across various layers?

\begin{figure}

\centering
\begin{subfigure}[b]{0.90\textwidth}
   \includegraphics[width=1\linewidth]{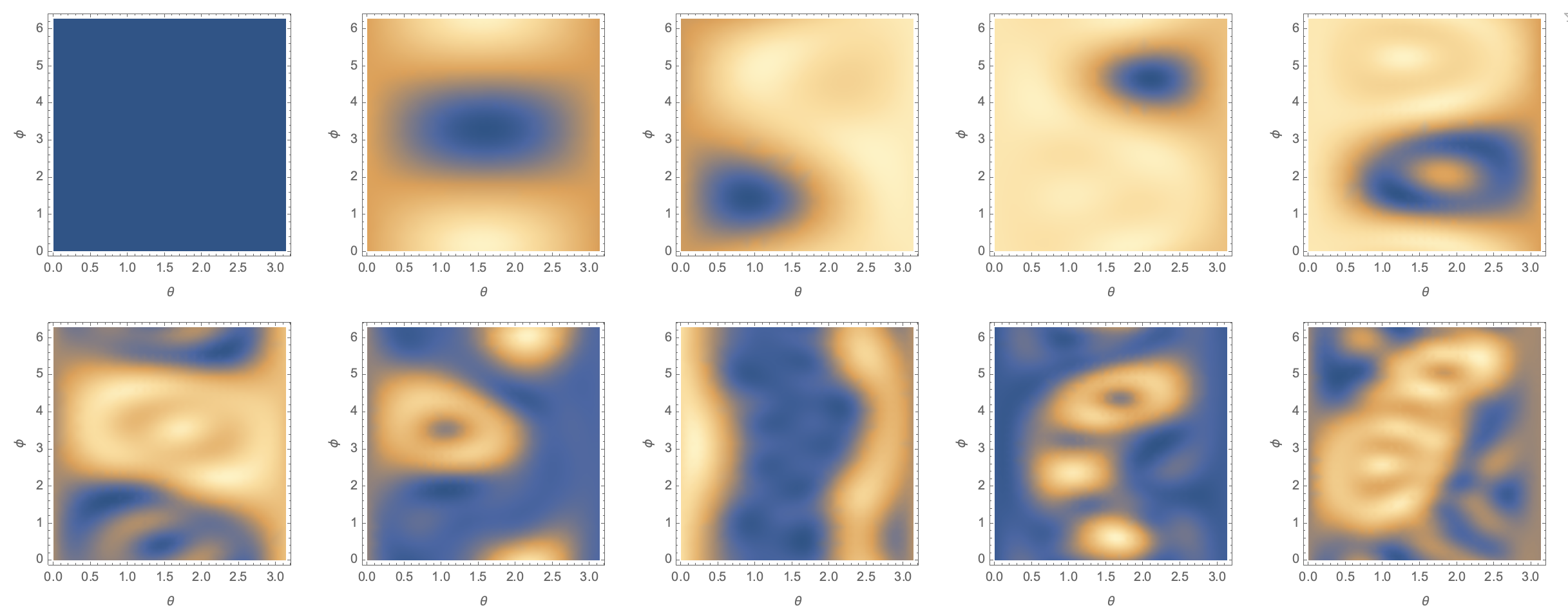}
   \caption{} 
\end{subfigure}

\begin{subfigure}[b]{0.90\textwidth}
   \includegraphics[width=1\linewidth]{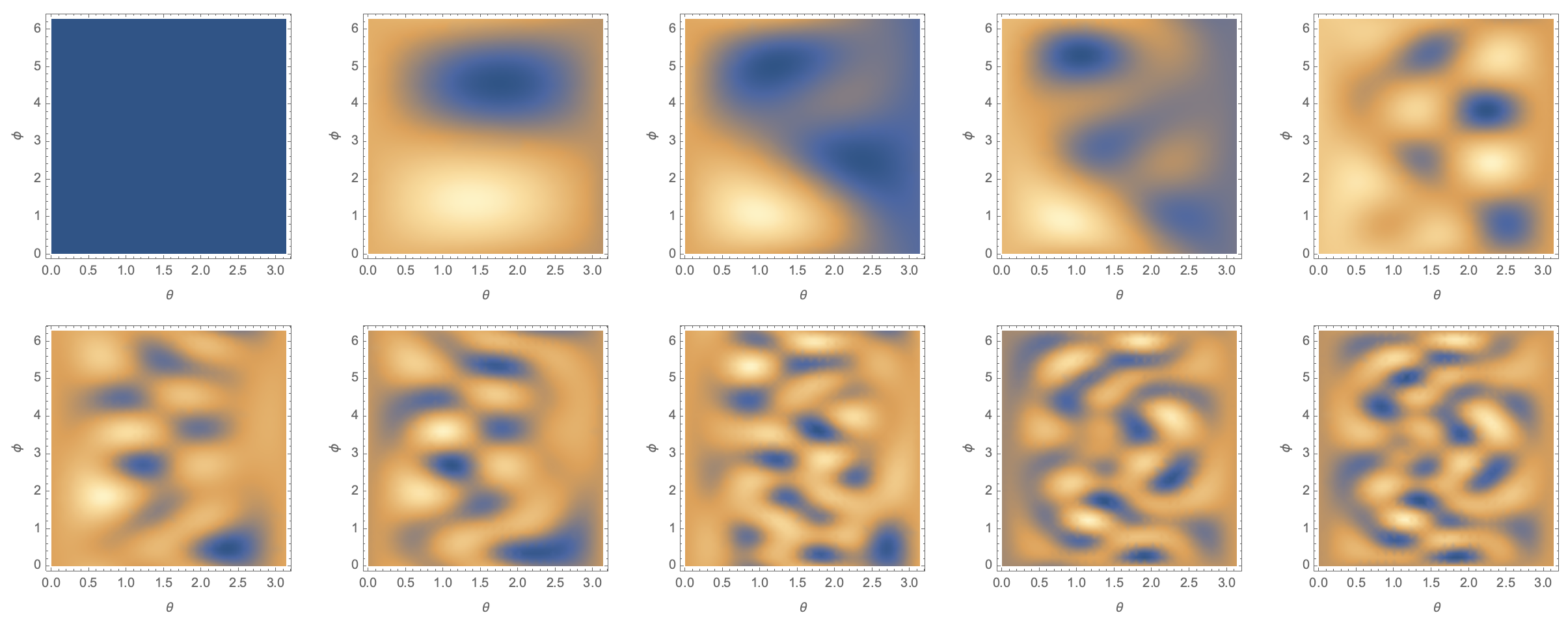}
   \caption{}
\end{subfigure}
\caption{Plots of $\Phi(\theta,\varphi)$ in a nonuniformly ordered phase with $a=1, b= - 3, c = 5$ and $N=10$ and the radial term missing (a) or present (b). We can see that including the radial term caused alignment across consecutive layers. Figure taken from \cite{Kovacik:2023zab}.}

\label{fieldtheory}
\end{figure}

To answer the first question, we have set a simulation in a regime where we expected the field theory to be in the nonuniformly ordered phase and saved a single configuration of the fields on the fuzzy onion after a sufficient number of steps after the thermalisation had been achieved. In the first simulation, the radial part of the kinetic term has been removed, and we can observe a range of wild patterns on each layer, completely independent from each other, see Figure \ref{fieldtheory}. This is an obvious outcome as, without the radial term, the simulation is basically a simulation of distinct fuzzy spheres. This serves as a benchmark of what noninteraction between the layers looks like. 

Then, we repeated the same simulation but with the radial term included. The alignment is now obvious; see Figure \ref{fieldtheory}. In principle, we could have calculated correlation coefficients to be more precise, but in this case, we believe the plots are sufficient to prove the point.

To answer the second question, we ran a simulation with the largest matrix of size $N=8$, the radial term included and values $b = -3.07,c = 12.8$. The eigenvalues were measured after each $10^3$ steps taken. The simulation was initiated in a uniformly ordered phase; that is, for each of the layers, all of the eigenvalues were in the same potential well. The simulations have been pushed strongly, meaning the acceptance rate was close to $50\%$, and the phase seemed stable. However, it turned out to be actually meta-stable only as around the step $220 \times 10^3$, first one and then shortly after, half of the eigenvalues on the outermost layer moved to the other potential well, and the field moved to the nonuniformly ordered phase. This effect then cascaded down and shifted all but two innermost layers to the same phase. We expect that in a properly ergodic simulation, actually, even the second innermost layers would shift and then the innermost layer would move between two layers. 

\begin{figure}
    \centering
    \includegraphics[width=1\linewidth]{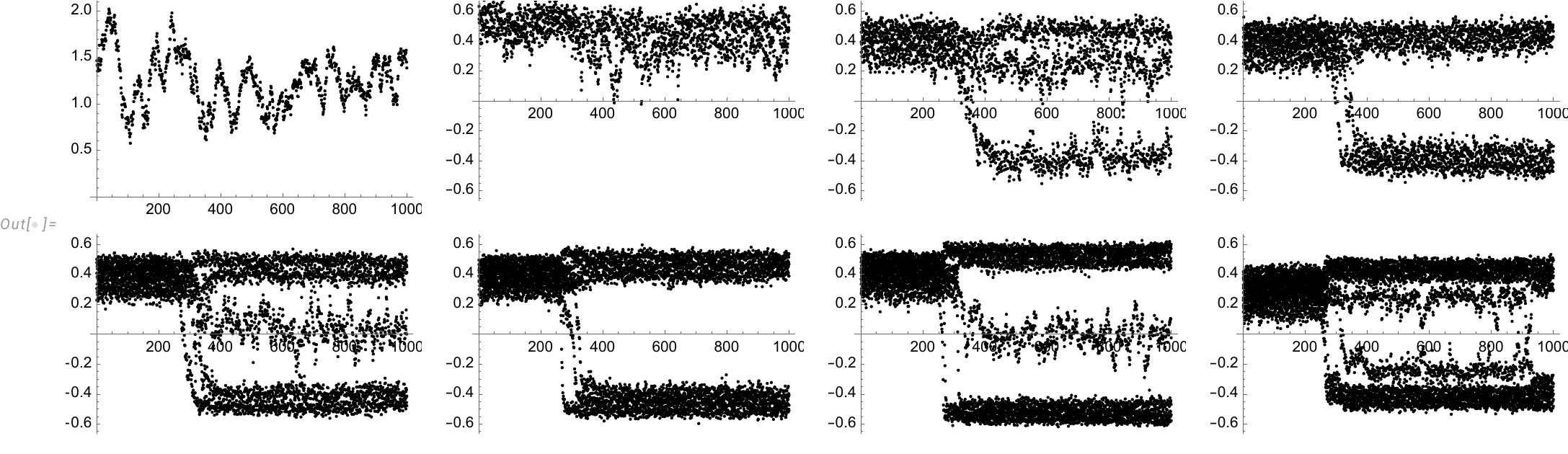}
    \caption{HMC trajectories for eigenvalues of $\Phi^{(N)}$ on each of the layers. We can see that while all of them started in the same phase, the change first happened at the outermost layer $\Phi^{(8)}$ and moved down to the $\Phi^{(3)}$ layer.}
    \label{fuzzy_phases}
\end{figure}

\section{Conclusion}

We have performed an initial study of the model of the scalar field theory on the fuzzy onion space. So far, the model behaves as expected. As we have shown in \cite{Kovacik:2023zab}, the model reproduces the results of the hydrogen atom problem in noncommutative space and can be used to study some classical problems, such as the heat transfer. 

Here, we have discussed in more detail the behaviour of the scalar field theory. Again, the model seems to be working as intended. The radial term allows the interaction of neighbouring layers, and it causes not only the alignment of the fields but also a rather abrupt alignment of the phases. 

Obviously, more detailed studies are warranted. In the first regard, we are interested in the large $N$ limit of the model. Also, it would be interesting to see if fine-tuning the coefficients $(b,c)$ would allow for the coexistence of multiple phases. Another line of research is to verify various ways of connecting the potential terms across various layers, either in the smearing or string formalism mentioned earlier in this report. 

\acknowledgments
This research was supported by the VEGA 1/0703/20 grant and the MUNI Award for Science and Humanities funded by the Grant Agency of Masaryk University.

\end{document}